\documentclass[twocolumn,showpacs,preprintnumbers,amsmath,amssymb,aps,prc,superscriptaddress,reprint,floatfix]{revtex4-2}
\usepackage{graphicx}
\usepackage{dcolumn}
\usepackage{bm}
\usepackage[utf8]{inputenc}
\usepackage{multirow}
\usepackage{tabularx}
\usepackage{microtype}
\microtypesetup{protrusion=true,expansion=true,tracking=true,kerning=true}
\microtypecontext{spacing=nonfrench}
\usepackage{hyperref}
\hypersetup{unicode=false,colorlinks=true,breaklinks=true,citecolor=blue,urlcolor=blue,linkcolor=blue}
\begin{document}

\title{Extraction of the non-spin- and spin-transfer isovector responses via the $^{12}\mathrm{C}(^{10}\mathrm{Be},{}^{10}\mathrm{B}+\gamma)^{12}\mathrm{B}$ reaction}

\author{Sk~M.~Ali}
\email{mustak@frib.msu.edu}
\thanks{The first two authors contributed equally to this work.}
\affiliation{Facility for Rare Isotope Beams, Michigan State University, East Lansing, Michigan 48824, USA}
\author{C.~Maher}

\email{maher@frib.msu.edu}
\affiliation{Facility for Rare Isotope Beams, Michigan State University, East Lansing, Michigan 48824, USA}
\affiliation{Department of Physics and Astronomy, Michigan State University, East Lansing, Michigan 48824, USA}
\author{R.~G.~T.~Zegers}
\email{zegers@frib.msu.edu}
\affiliation{Facility for Rare Isotope Beams, Michigan State University, East Lansing, Michigan 48824, USA}

\affiliation{Department of Physics and Astronomy, Michigan State University, East Lansing, Michigan 48824, USA}
\author{M.~Scott}
\affiliation{Facility for Rare Isotope Beams, Michigan State University, East Lansing, Michigan 48824, USA}

\affiliation{Department of Physics and Astronomy, Michigan State University, East Lansing, Michigan 48824, USA}

\author{D.~Bazin}
\affiliation{Facility for Rare Isotope Beams, Michigan State University, East Lansing, Michigan 48824, USA}
\affiliation{Department of Physics and Astronomy, Michigan State University, East Lansing, Michigan 48824, USA}
\author{M.~Bowry}
\thanks{Current Address: Nuclear Physics Research Center, School of Computing Engineering and Physical Sciences, University of the West of Scotland, Paisley PA1 2BE, United Kingdom}
\affiliation{Facility for Rare Isotope Beams, Michigan State University, East Lansing, Michigan 48824, USA}
\author{B.~A.~Brown}
\affiliation{Facility for Rare Isotope Beams, Michigan State University, East Lansing, Michigan 48824, USA}

\affiliation{Department of Physics and Astronomy, Michigan State University, East Lansing, Michigan 48824, USA}
\author{C.~M.~Campbell}
\affiliation{Nuclear Science Division, Lawrence Berkeley National Laboratory, Berkeley, California 94720, USA}
\author{A.~Gade}
\affiliation{Facility for Rare Isotope Beams, Michigan State University, East Lansing, Michigan 48824, USA}
\affiliation{Department of Physics and Astronomy, Michigan State University, East Lansing, Michigan 48824, USA}
\author{S.~Galès}
\thanks{Deceased.}
\affiliation{IPN Orsay, CNRS-IN2P3, Universit\'e Paris-Sud, Universit\'e Paris-Saclay, 91406 Orsay Cedex, France}
\affiliation{Horia Hulubei National Institute of Physics and Nuclear Engineering, P.O. Box MG6, Bucharest, Romania}
\author{U.~Garg}
\affiliation{Department of Physics, University of Notre Dame, Notre Dame, Indiana 46556, USA}
\author{M.~N.~Harakeh}

\affiliation{Energy and Sustainability Research Institute Groningen (ESRIG), University of Groningen, Groningen, 9747 AA, Netherlands} 
\author{E.~Kwan}
\affiliation{Facility for Rare Isotope Beams, Michigan State University, East Lansing, Michigan 48824, USA}
\author{C.~Langer}
\thanks{Current Address: University of Applied Sciences Aachen, Campus Jülich, 52428 Jülich, Germany}
\affiliation{Facility for Rare Isotope Beams, Michigan State University, East Lansing, Michigan 48824, USA}

\author{C.~Loelius}

\affiliation{Department of Physics and Astronomy, Michigan State University, East Lansing, Michigan 48824, USA}

\author{C.~Morse}
\affiliation{Facility for Rare Isotope Beams, Michigan State University, East Lansing, Michigan 48824, USA}
\affiliation{Department of Physics and Astronomy, Michigan State University, East Lansing, Michigan 48824, USA}
\author{S.~Noji}
\affiliation{Facility for Rare Isotope Beams, Michigan State University, East Lansing, Michigan 48824, USA}

\author{T.~Redpath}
\thanks{Current Address: Department of Chemistry, Virginia State University, Petersburg 23806, USA}
\affiliation{Department of Physics, Central Michigan University, Mount Pleasant, Michigan 48859, USA}

\author{H.~Sakai}
\affiliation{RIKEN Nishina Center, Hirosawa, Wako, Saitama 351-0198, Japan}

\author{M.~Sasano}
\affiliation{RIKEN Nishina Center, Hirosawa, Wako, Saitama 351-0198, Japan}
\author{C.~Sullivan}
\affiliation{Facility for Rare Isotope Beams, Michigan State University, East Lansing, Michigan 48824, USA}

\affiliation{Department of Physics and Astronomy, Michigan State University, East Lansing, Michigan 48824, USA}

\author{D.~Weisshaar}
\affiliation{Facility for Rare Isotope Beams, Michigan State University, East Lansing, Michigan 48824, USA}

\date{\today}
\begin{abstract}  
The isovector response in $^{12}$B was investigated via the $^{12}$C($^{10}\mathrm{Be}$,$^{10}\mathrm{B}$+$\gamma$)$^{12}$B$^\ast$ reaction at $100 A \, \mathrm{MeV}$. By utilizing the $\gamma$-decay properties of the 1.74 MeV $0^{+}$ and 0.718 MeV $1^{+}$ states in $^{10}\mathrm{B}$, the separate extraction of the non-spin-transfer ($\Delta S=0$) and spin-transfer ($\Delta S=1$) isovector responses up to an excitation energy of 50 MeV in $^{12}$B in a single measurement is demonstrated.
The experimental setup employed the S800 spectrometer to detect and analyze the $^{10}\mathrm{B}$ ejectiles and the Gamma-Ray Energy Tracking In-beam Nuclear Array (GRETINA) for obtaining the Doppler-reconstructed spectrum for $\gamma$-rays emitted in-flight by $^{10}\mathrm{B}$.
A $^{12}$C foil was placed at the pivot point of the spectrograph. The $^{12}$B reaction product was not detected.
Contributions from transitions associated with the transfer of different units of angular momentum in the non-spin- and spin-transfer responses were analyzed using a multipole decomposition analysis. The extracted non-spin-dipole ($\Delta S=0$, $\Delta L=1$) and spin-dipole ($\Delta S=1$, $\Delta L=1$) responses were found to be consistent with available data from other charge-exchange probes, validating the non-spin- and spin-transfer filters used. While statistical uncertainties and experimental resolutions were relatively large due to the modest intensity of the $^{10}\mathrm{Be}$ secondary beam, the results show that, with the much higher intensities that will be available at new rare-isotope beam facilities, the ($^{10}\mathrm{Be}$,$^{10}\mathrm{B}$+$\gamma$) reaction and its $\Delta T_{z}=-1$ partner, the ($^{10}$C,$^{10}\mathrm{B}$+$\gamma$) reaction, are powerful tools for elucidating the isovector non-spin- and spin-transfer responses in nuclei.     

\end{abstract}
 
\maketitle

\section{Introduction}
The study of spin-isospin excitations provides valuable insight into the isovector properties of nuclei and the macroscopic properties of nuclear matter~\cite{OST92,Harakeh2001}, with significant implications for astrophysics and neutrino physics~\cite{Langanke2003, Langanke2021}. Experimentally, charge-exchange (CE) reactions at intermediate energies ($E \gtrsim 100 A \, \mathrm{MeV}$) have been extensively used to investigate the isovector response in nuclei including giant resonances~\cite{Harakeh2001,ICHIMURA2006,Zegers2023}. At these beam energies, multistep contributions to the reaction mechanism are minimal, allowing the CE reaction to be treated as a direct, one-step process. Unlike $\beta$-decay measurements, CE reactions are not limited by the $Q$ value and thus can be used to test theoretical models up to high excitation energies.
A wide variety of CE probes, ranging from nucleonic probes $(p,n)$/$(n,p)$ to light-ion probes such as $(d,{}^2\mathrm{He})$ and $(t,{}^3\mathrm{He})$/$(^3\mathrm{He},t)$, have been employed to excite isovector transitions in both the $\beta^-$ ($\Delta T_z=-1$) and $\beta^+$ ($\Delta T_z=+1$) directions~\cite{ICHIMURA2006,Fujita2011549,FRE18,Zegers2023}.
Heavier composite probes ($A>3$) with stable and unstable beams have also been successfully utilized as they provide new ways to isolate specific giant resonances. These include the $(^6\mathrm{Li},{}^6\mathrm{He})$~\cite{UENO1999} reaction, the $(^7\mathrm{Li},{}^7\mathrm{Be}+\gamma)$ reaction~\cite{Nakayama1991,Winfield1996,NAKAYAMA1998,ANNAKKAGE1999,Zegers2010, Meharchand2012}, the $(^{12}\mathrm{C},{}^{12}\mathrm{N})$ and $(^{12}\mathrm{C},{}^{12}\mathrm{B})$ reactions~\cite{Aanantaraman1991,Ichihara1994,ICHIHARA1994PLB,ICHIHARA1995}, and the ($^{13}$C,$^{13}$N) reaction~\cite{BERAT1993455,Ichihara2002}. The advent of rare-isotope beam facilities enabled the use of
the $(p,n)$ and $(d,{}^2\mathrm{He})$ CE probes to study rare isotopes in inverse kinematics~\cite{Sasano2011,Sasano2012,Yasuda2018,Giraud2023,Rahman2024}.
In contrast, unstable heavy-ion CE probes have also been utilized in forward kinematics, such as the $(^{13}\mathrm{N},{}^{13}\mathrm{C})$~\cite{Steiner1996}, the $(^{12}\mathrm{N},{}^{12}\mathrm{C})$~\cite{Noji2018}, $(^{10}\mathrm{C},{}^{10}\mathrm{B}+\gamma)$~\cite{Sasamoto2012,Sasamoto2012a,Uesaka2012}, and $(^{10}\mathrm{Be},{}^{10}\mathrm{B}+\gamma)$~\cite{Scott2017} reactions.
The reason for the use of different probes is that they have different experimental properties, advantages, and varying selectivity for the type of excited isovector transitions~\cite{Zegers2023}. For example, the $(d,{}^2\mathrm{He})$, $(^{6}\mathrm{Li},{}^6\mathrm{He})$, $(^{12}\mathrm{C},{}^{12}\mathrm{N})$, and $(^{12}\mathrm{C},{}^{12}\mathrm{B})$, and $(^{12}\mathrm{N},{}^{12}\mathrm{C})$ probes are selective for excitations associated with spin-transfer ($\Delta S=1$). The $(^{13}\mathrm{C},{}^{13}\mathrm{N})$ and $(^{13}\mathrm{N},{}^{13}\mathrm{C})$ probes allow for both spin-transfer and non-spin-transfer ($\Delta S=0$) excitations, the latter of which is favored due to the large Fermi matrix elements ($M_\mathrm{F}/M_\mathrm{GT} \approx 5$)~\cite{BERAT1993455}.

Only a few CE probes can be used to isolate $\Delta S=0$ and $\Delta S=1$ responses in a single measurement. A probe with such an ability eliminates systematic uncertainties in the comparison of the two responses when two different probes and experiments must be utilized.  The $(^7\mathrm{Li},{}^7\mathrm{Be}+\gamma)$ reaction allows for the separation of the spin- and non-spin-transfer isovector transitions from the same measurement. In this reaction, by detecting the $^7\mathrm{Be}$ ion in coincidence with the 429~keV $\gamma$-ray emitted following the $^7\mathrm{Li}(3/2^-,\mathrm{g.s.}) \rightarrow {}^7\mathrm{Be}(1/2^-,0.429~\mathrm{MeV})$ transition, a clean $\Delta S=1$ filter is obtained~\cite{Nakayama1991,Winfield1996,NAKAYAMA1998,ANNAKKAGE1999,Zegers2010}. The transition to the $^7\mathrm{Be}$ ground state, $^7\mathrm{Li}(3/2^-,\mathrm{g.s.}) \rightarrow {}^7\mathrm{Be}(3/2^-,\mathrm{g.s.})$ involves both $\Delta S = 1$ and $\Delta S = 0$ transitions. The $\Delta S=0$ response can be extracted by subtracting the $\Delta S=1$ contribution obtained from the above-mentioned transition to the $1/2^-$ state. However, performing this subtraction requires a detailed understanding of the relative strengths of the $\tau$, $\sigma\tau$ and tensor-$\tau$ components of the nucleon-nucleon interaction that mediate the CE reaction, as well as the wave functions of the relevant states in $^7\mathrm{Li}$ and $^7\mathrm{Be}$. The necessary theoretical estimation introduces a systematic uncertainty in the extracted $\Delta S=0$ response.

To extract the $\Delta S=0$ and $\Delta S=1$ responses within a single  measurement, the $(^{10}\mathrm{Be},{}^{10}\mathrm{B}+\gamma)$ reaction is a promising alternative. This reaction probe was first developed to isolate the $\Delta S=0$ isovector giant monopole resonance (IVGMR) in $^{28}$Si~\cite{Scott2017}. This probe has a $\beta^-$ counterpart, the $(^{10}\mathrm{C},{}^{10}\mathrm{B}+\gamma)$ reaction, which has also been studied~\cite{Sasamoto2012,Sasamoto2012a,Uesaka2012}. While those experiments primarily aimed at isolating the $\Delta S=0$ response by selecting the reaction channel leading to the $J^\pi = 0^+$, $T=1$ state at 1.74~MeV in $^{10}\mathrm{B}$ via $\gamma$-ray coincidence, a $\Delta S=1$ filter was also obtained by identifying the reaction channel leading to the $J^\pi = 1^+$, $T=0$ state at 0.718~MeV in $^{10}\mathrm{B}$. In both cases, to extract uncontaminated $\Delta S=0$ and $\Delta S=1$ responses, feeding from the higher-lying states in $^{10}\mathrm{B}$ must be subtracted. These subtractions solely rely on the well-studied decay branching ratios and are independent from the reaction mechanism. Consequently, these probes are attractive options for the simultaneous extraction of the $\Delta S=0$ and $\Delta S=1$ responses within a single measurement.

In the present work, the $^{12}\mathrm{C}(^{10}\mathrm{Be},{}^{10}\mathrm{B}+\gamma)$ reaction at $100 A~\mathrm{MeV}$ was employed to simultaneously extract the $\Delta S=0$ and $\Delta S=1$ isovector excitations to $^{12}\mathrm{B}$ with the goal to assess the viability of the approach.
The $^{12}$B is produced in the CE reactions that take place in the $^{12}$C foil at the target location.
The experimental data were taken in the same experimental campaign to study the $\Delta S=0$ excitations from $^{28}\mathrm{Si}$~\cite{Scott2017}. The isovector transitions in the $^{12}\mathrm{C} \rightarrow {}^{12}\mathrm{B}$ system have been well studied using various CE reactions on $^{12}\mathrm{C}$ such as the $(n,p)$~\cite{OLSSON1993,Yang1993}, $(d,{}^2\mathrm{He})$~\cite{OKAMURA1994,DEHUU2007}, $(^7\mathrm{Li},{}^7\mathrm{Be})$~\cite{SAKUTA2006,ANNAKKAGE1999,Nakayama1991,NAKAYAMA1998}, $(^{12}\mathrm{C},{}^{12}\mathrm{N})$ and $(^{13}\mathrm{C},{}^{13}\mathrm{N})$~\cite{Ichihara1994,ICHIHARA1994PLB,ICHIHARA1995} reactions. Thus, $^{12}\mathrm{B}$ serves as an excellent case to test the effectiveness of the $(^{10}\mathrm{Be},{}^{10}\mathrm{B})$ CE reaction in separating the $\Delta S=0$ and $\Delta S=1$ components. This is important for planning future experiments, which will benefit from increased $^{10}\mathrm{Be}$ beam intensities compared to the experiment presented here.

\section{Experiment}
The $^{12}\mathrm{C}(^{10}\mathrm{Be},{}^{10}\mathrm{B}+\gamma)$ experiment was performed at the Coupled Cyclotron Facility of the National Superconducting Cyclotron Laboratory (NSCL). The $^{10}\mathrm{Be}$ beam was produced by impinging a $150~\mathrm{pnA}$, $120 A~\mathrm{MeV}$ $^{18}\mathrm{O}$ beam on a 1316~mg/cm$^{2}$-thick Be target at the entrance of the A1900 fragment separator~\cite{MORRISSEY200390}. After purification in the A1900, a secondary $^{10}\mathrm{Be}$ beam of $100 A~\mathrm{MeV}$ was impinged upon a 56.33~mg/cm$^2$-thick $^\textrm{nat}$C target (98.88\% $^{12}\mathrm{C}$) placed at the pivot point of the S800 spectrograph~\cite{S800}. The $^{10}\mathrm{Be}$ beam, with a momentum spread of $\lvert \Delta p/p \rvert \leqslant 0.25\%$, was transported to the S800 target station in the dispersion-matched ion-optical mode~\cite{FUJITA200217}, enabling the reconstruction of the excitation-energy spectra with better resolution than the energy spread in the beam. An incoming beam rate of $\approx$7~MHz and purity of 98\% were achieved for the $^{10}\mathrm{Be}$ beam at the target location of the S800 spectrograph.

The reaction products, including $^{10}\mathrm{B}$ from the $(^{10}\mathrm{Be},{}^{10}\mathrm{B})$ reaction were magnetically analyzed for momentum and detected in the S800 focal-plane detector system consisting of two cathode readout drift chambers (CRDCs) installed 1 m apart, an ionization chamber, and a 5 mm-thick plastic scintillator~\cite{YURKON1999291}.
The CRDCs measured the positions and angles of the reaction products. The two-dimensional positions from each CRDC were calibrated by placing thick tungsten masks with holes and slits at known positions in front of it, which was then irradiated with the $^{10}\mathrm{Be}$ beam.  
The ionization chamber downstream of the second CRDC measured the energy loss of the reaction products.
The rearmost plastic scintillator provided the trigger for the data-acquisition system as well as the timing information. The time-of-flight (TOF) of the reaction products was deduced from the scintillator timing relative to the radio frequency (RF) signal of the cyclotrons.
To separate the $^{10}\mathrm{B}$ ions from the other reaction products, the energy loss in the  ionization chamber ($\Delta E$) was plotted against the TOF of the ejectiles, and gated on the residues of interest.
The energy, outgoing angles in the dispersive and non-dispersive directions, and the position in the non-dispersive direction at the target location were deduced from the positions and angles in the focal plane in a ray-tracing procedure using an inverse map calculated with the ion-optical code COSY Infinity~\cite{cosy}. 
The differential cross sections were calculated across the excitation-energy range of $0\leqslant \textrm{E}_x \leqslant 50$ MeV and the scattering-angle range of $0^\circ \leqslant \theta_\textrm{c.m.} \leqslant 3^\circ$ in a missing-mass calculation. The experimental excitation-energy and angular resolutions (FWHM) were 2.0~MeV and $0.5^\circ$, respectively. The excitation-energy resolution was obtained by using the $^{12}$C(0$^+$,g.s.)$\rightarrow$$^{12}$B(1$^+$,g.s.) peak (detailed below). The resolution is due to the intrinsic momentum resolution of the measurement and the difference in energy loss between $^{10}$Be and $^{10}$B in the $^{12}$C target foil.
The resolution of the scattering angle was obtained through a measurement of the unreacted beam in the focal plane of the S800 spectrometer.

To isolate the $\Delta S=0$ and $\Delta S=1$ isovector excitations, it is essential to detect the $\gamma$-rays emitted in-flight from the de-excitation of the $^{10}\mathrm{B}$ ions with a high signal-to-noise ratio. Therefore, the Gamma-Ray Energy Tracking In-beam Nuclear Array (GRETINA)~\cite{WEISSHAAR2017187,GRETINA} was used in the present experiment. Seven Quad modules of GRETINA were placed at $90^\circ$ relative to the beam axis around the location of the carbon target. The $\gamma$-ray detection efficiency, determined from measurements with standard calibration sources, ranged from 0.114 for $E_{\gamma}=200$ keV to 0.034 for $E_{\gamma}=2000$ keV.

As the $^{10}\mathrm{B}$ ejectiles were traveling at ${\approx}40\%$ of the speed of light during the $\gamma$-decay, Doppler reconstruction was necessary to determine the energy of the $\gamma$-rays in the rest frame of $^{10}\mathrm{B}$. The Doppler-corrected energy resolution was affected by the uncertainty in the determination of emission angles of the $\gamma$-rays, which was aggravated by a large beam-spot size (about ${\pm}2.5~\mathrm{cm}$ in the dispersive direction for $\lvert \Delta p/p \rvert \leqslant 0.25\%$) resulting from the dispersion-matching beam transport. Placing the GRETINA detectors at $90^\circ$ with respect to the beam axis helped mitigate such effects.

\section{Data Analysis}
\subsection{Non-spin- and spin-transfer excitations}
\begin{figure}[tbp]
\includegraphics[width=0.5\textwidth,clip]{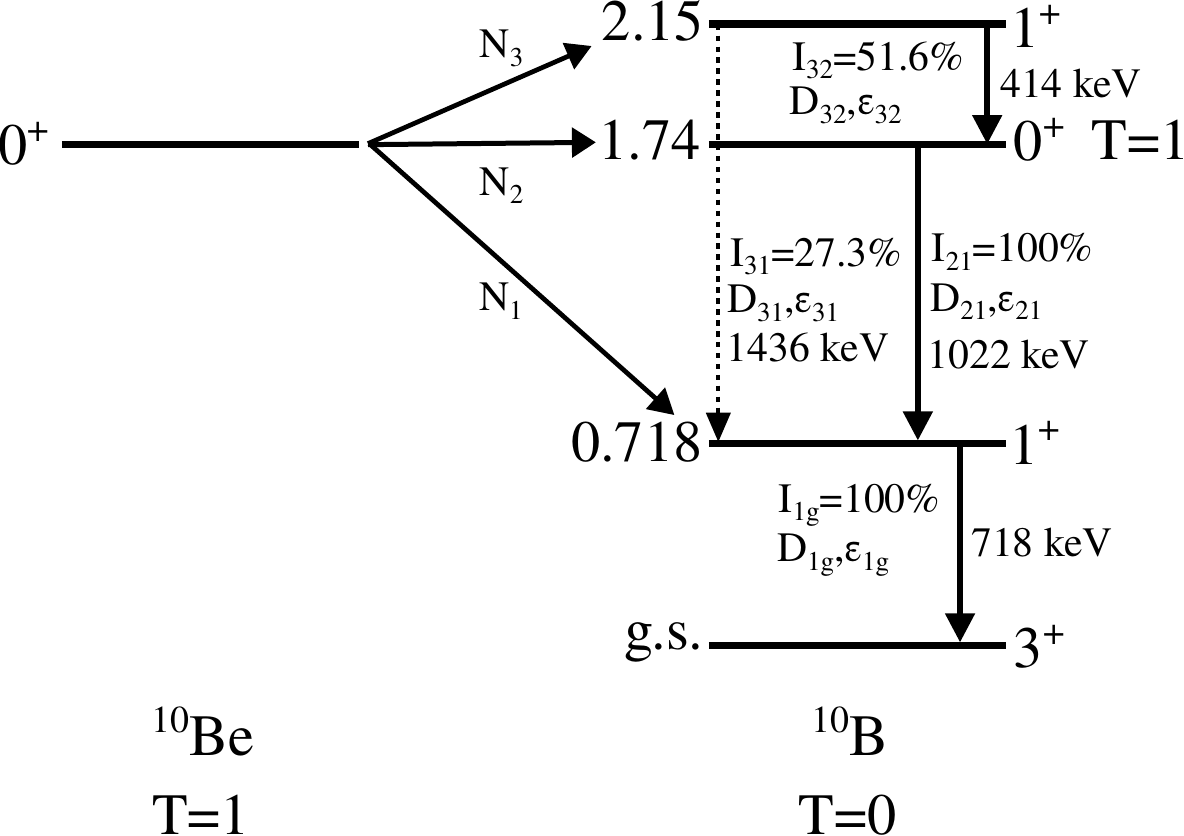}
\caption{The  $^{10}\mathrm{B}$  level scheme relevant for the ($^{10}\mathrm{Be}$,$^{10}\mathrm{B}$+$\gamma$) reaction probe to isolate the $\Delta S = 0$ and $\Delta S=1$ transitions.} 
\label{fig:level scheme}
\end{figure}

Events in which $^{10}\mathrm{B}$ was detected in the S800 focal plane, coincident with specific $\gamma$-rays emitted during the in-flight de-excitation of $^{10}\mathrm{B}$, enabled the identification of the isovector non-spin-transfer ($\Delta S=0$, $\Delta T=1$) and spin-transfer ($\Delta S=1$, $\Delta T=1$) excitations. The relevant levels and their properties needed for the subsequent analysis are shown in Fig.~\ref{fig:level scheme}. The number of transitions to the first-excited state at 0.718~MeV, the second-excited state at 1.74~MeV, and the third-excited state at 2.15~MeV are denoted as $N_1$, $N_2$, and $N_3$, respectively. The numbers of observed $\gamma$-rays of interest are labeled as $D_{if}$, where $i$ and $f$ refer to either the ground state ($g$) or an excited state (1, 2, or 3). The relevant branching percentages ($I_{if}$) are shown in the figure, and the $\gamma$ detection efficiencies for each of the decays are referred to as $\epsilon_{if}$.

The Fermi transition strength from the $^{10}\mathrm{Be}$ ground state to its isobaric analog state (IAS) at 1.74~MeV in $^{10}\mathrm{B}$ is $B(\mathrm{F})=2$. The non-spin-transfer filter is achieved by selecting events in which the 1022~keV $\gamma$-ray is emitted from the 1.74~MeV $0^+$ state to the 0.718~MeV $1^+$ state with a branching percentage $I_{21}=100\%$. To obtain a clean filter, feeding from higher-lying states must be subtracted. The level scheme of $^{10}\mathrm{B}$ in Fig.~\ref{fig:level scheme} indicates that the 2.15~MeV $1^+$ state de-excites by emitting a 414~keV $\gamma$-ray into the 1.74~MeV $0^+$ state with a probability $I_{32} = 51.6\%$, contaminating the $\Delta S=0$ filter. In addition, feeding from the 3.59~MeV $2^+$ state is also possible, but the population of this $2^+$ state through the CE reaction was very weak and thus could not be observed in the experiment. Therefore, this feeding was ignored in the further analysis. Consequently, the number of transitions $N_{2}$ to the 1.74~MeV state was determined from
\begin{align}
    N_{{2}}=\frac{D_{21}}{\epsilon_{21}}-\frac{D_{32}}{\epsilon_{32}}.
\end{align}

Based on the measured $\log \mathit{ft}$ value~\cite{TILLEY2004155} for the analog $\beta^-$ decay of $^{10}$C to the $^{10}\mathrm{B}$ $1^+$ excited state at 0.718~MeV and on the isospin symmetry, the Gamow-Teller (GT) transition strength from $^{10}\mathrm{Be}$ to the same 0.718~MeV state in $^{10}\mathrm{B}$ is $B(\mathrm{GT}) = 3.51$. This transition, which only decays through a 718~keV $\gamma$-ray ($I_{1g}=100\%)$, was used to enable the spin-transfer filter. Here, feeding from the $0^+$ state at 1.74~MeV with a probability of 100\% ($I_{21}$) and that from the $1^{+}$ state at 2.15~MeV with a 27.3\% probability ($I_{31}$) needed to be subtracted. Considering that $I_{32}$ is larger than $I_{31}$, and the detection efficiency and hence the signal-to-background ratio for the 414~keV $\gamma$-ray is higher than that for the 1436~keV $\gamma$-ray, the spectrum to be subtracted was generated by gating on the 414~keV $\gamma$-rays, with the difference in branching ratios compensated for. We note that the excitation of the 2.15~MeV state is also associated with the transfer of spin. However, since the associated GT strength is very small (only an upper limit of 0.007 has been established \cite{FUJIKAWA19996}), unlike the strong transition to the $1^{+}$ state at 0.718~MeV, the CE reaction to this 2.15~MeV state is likely to be complicated \cite{PhysRevC.74.024309} due to interference between $\Delta L=0$ and $\Delta L=2$ amplitudes mediated by the tensor-$\tau$ component of the nucleon-nucleon interaction. Given these complexities, transitions to the 2.15~MeV were not included in the spin-transfer filter to ensure its robustness in the present study. Consequently, the number of transitions $N_{1}$ to the 0.718~MeV state, which serves as the spin-transfer filter, was determined from
\begin{align}
    N_{{1}}=\frac{D_{1g}}{\epsilon_{1g}}-\frac{D_{21}}{\epsilon_{21}}-\frac{D_{31}}{\epsilon_{31}}=    
    \frac{D_{1g}}{\epsilon_{1g}}-\frac{D_{21}}{\epsilon_{21}}-\frac{D_{32}}{\epsilon_{32}}\frac{I_{31}}{I_{32}}.
\end{align}

\begin{figure}[tbp]
\includegraphics[width=0.5\textwidth,clip]{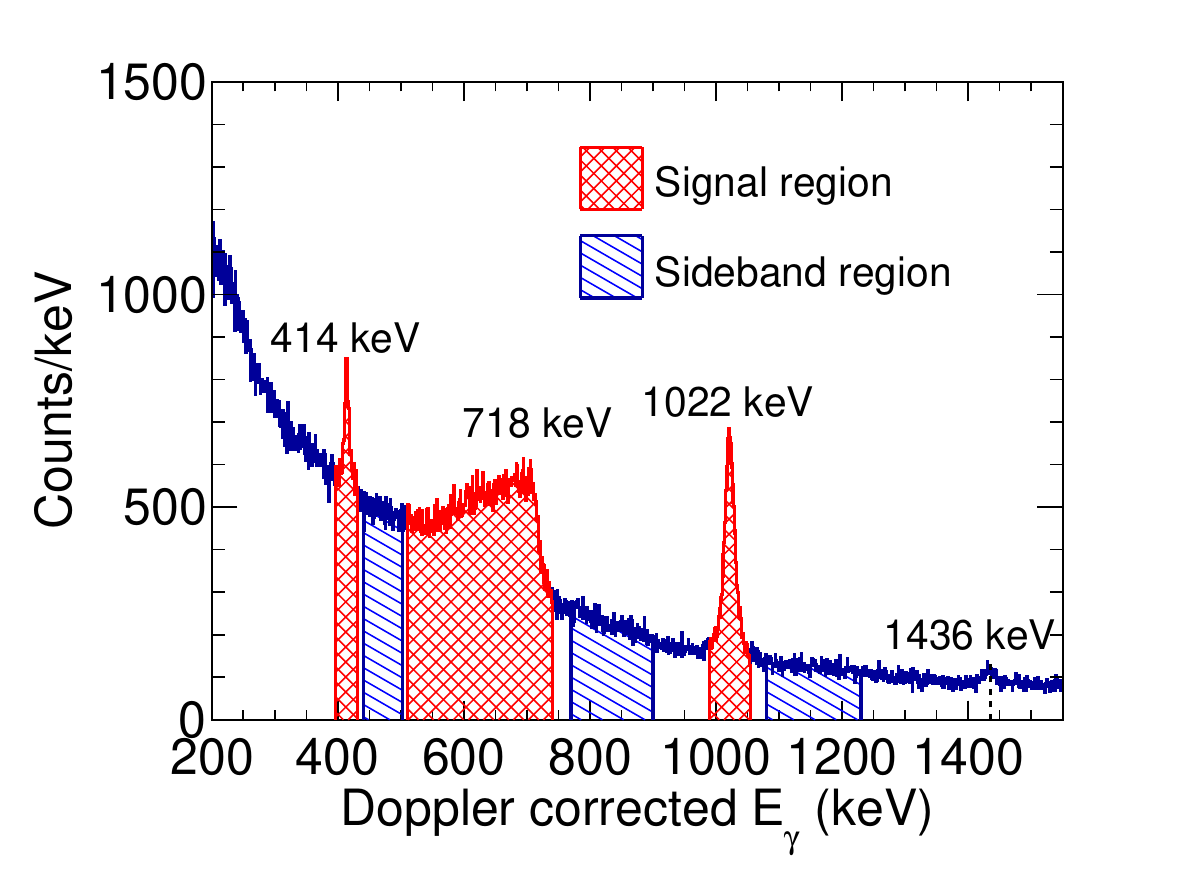}
\caption{Doppler-reconstructed $\gamma$-ray spectrum
from the $^{12}$C($^{10}\mathrm{Be}$,$^{10}\mathrm{B}$+$\gamma$)$^{12}$B reaction at $100 A \, \mathrm{MeV}$. The signal and sideband regions for the 414, 718 and 1022~keV $\gamma$-peaks are shown in red and blue, respectively.} \label{fig:gamma_spectrum}
\end{figure}
 
The Doppler-reconstructed $\gamma$-ray-energy spectrum is shown in Fig.~\ref{fig:gamma_spectrum}. Peaks are observed at 414~keV, ${\approx}718$~keV, 1022~keV, and 1436~keV, associated with the transitions identified in Fig.~\ref{fig:level scheme}. The detection efficiencies for the Doppler-corrected $\gamma$-rays were determined by averaging the efficiency for the $\gamma$-ray for each event, using its laboratory-frame energy and the calibrated efficiency curve. The average efficiency for the Doppler-corrected 414~keV $\gamma$-rays was determined to be $\epsilon_{32}=8.3\%$ and that for the 1022~keV $\gamma$-rays $\epsilon_{21}=5.1\%$.  

The $1^{+}$ state at 0.718~MeV has a half-life of 0.707~ns, corresponding to a distance traveled of about 9~cm after the target at a velocity of 40\% of the speed of light. Due to the incorrectly assumed $\gamma$-ray emission points for the events that decay at a distance from the target, the peak observed in the $\gamma$-ray spectrum of Fig.~\ref{fig:gamma_spectrum} was broadened, and events associated with this decay were observed down to an energy of 520~keV, corresponding to a distance traveled of about 36~cm after the target. The average detection efficiency for this transition was determined from the data by considering the $\gamma$-$\gamma$ coincidences recorded. Since the 1.74~MeV state always decays by emitting a 1022-keV $\gamma$-ray and goes to the 0.718~MeV state, which subsequently decays by emitting a 718~keV $\gamma$-ray, both of these $\gamma$-rays were always emitted from the 1.74~MeV state. Therefore, the efficiency for the 718~keV $\gamma$-ray can be calculated by comparing the numbers of the 718~keV $\gamma$-rays observed with and without gating on the 1022~keV peak. The background events under the 1022~keV peak that result in the detection of a 718~keV $\gamma$-ray must be subtracted by using a sideband analysis, as discussed in Sec.~\ref{sec:sideband}. In addition, the background events under the 718~keV $\gamma$-ray in the spectrum gated on the 1022~keV $\gamma$-ray must be subtracted, introducing a significant uncertainty due to the estimation of the background under this broad peak in the $\gamma$-$\gamma$ coincidence spectrum with limited statistics. Using this method, an efficiency for the detection of the 718~keV $\gamma$-ray of interest was determined to be $0.029\pm0.004$.

Alternatively, one can also take the ratio of the numbers of the 1022~keV $\gamma$-rays observed with and without gating on the 718~keV peak. This second method has a lower uncertainty, as the narrower 1022 keV peak facilitates the background estimation, even in the $\gamma$-$\gamma$ coincidence spectrum. Background events must be subtracted in a similar fashion to the first method, requiring an estimate of the background shape under the 718~keV peak, but this can be done in the $\gamma$-singles spectrum with high statistics, resulting in a smaller uncertainty than the analysis in the $\gamma$-$\gamma$ coincidence spectrum needed for the first method. Using the second method, the efficiency was determined for the detection of the 718~keV $\gamma$-line of $0.031\pm0.002$, which was used in the remainder of the analysis.       

\subsection{Sideband analysis}~\label{sec:sideband}
\begin{figure}[tbp]
\centering
\includegraphics[width=0.5\textwidth]{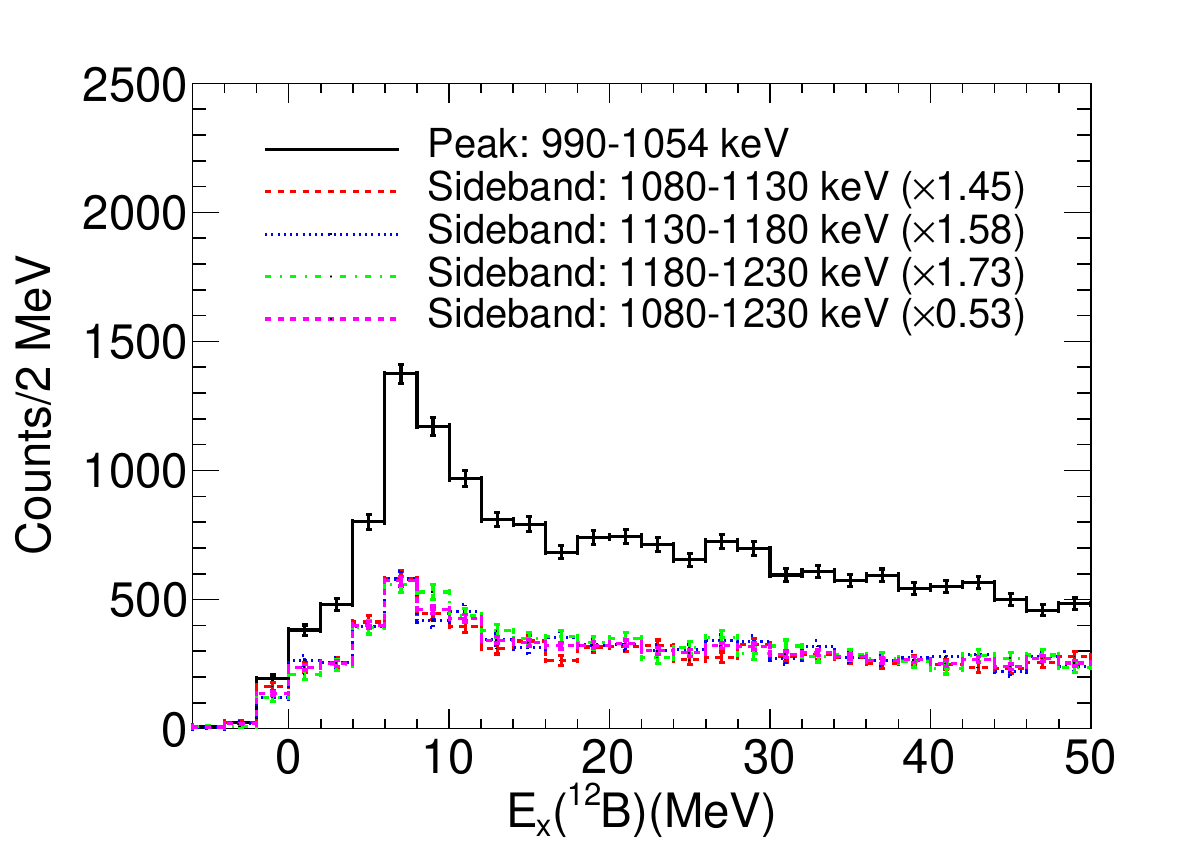}
\caption{Excitation-energy spectrum of $^{12}$B gated on the signal (solid black line) and different sideband regions (dashed and dotted lines in red, blue, green and magenta) of the 1022~keV $\gamma$-ray peak. The sideband spectra are scaled as described in the text.}  \label{fig:sideband}
\end{figure}

The high-resolution Doppler reconstruction afforded by using GRETINA allowed for the detection of $\gamma$-rays with good photo-peak signal-to-noise ratio, thereby reducing uncertainties in the background subtraction. Background contributions in the data were estimated through sideband studies in the $\gamma$-spectrum. The method involves selecting the signal and sideband regions in the $\gamma$-ray spectrum as illustrated by the red and blue shaded regions in Fig.~\ref{fig:gamma_spectrum}.  
The observed signal counts ($N_\textrm{signal}$) under the $\gamma$-peaks are obtained as
\begin{align}
    N_\textrm{signal}=N_\textrm{signal region}-kN_\textrm{sideband region},
\end{align}
where $N_\textrm{signal region}$ is the number of counts in the signal region under the peak of the $\gamma$-rays of interest and $N_\textrm{sideband region}$ is the number of counts in the sideband region. The sideband scaling factor ($k$) is the ratio of the number of counts in the background in the signal region to the number of counts in the sideband region. The sideband region was chosen to minimize the statistical uncertainties in $k$ while ensuring that the shape of the spectra associated with the sideband was independent of the sideband width. Figure~\ref{fig:sideband} presents a comparison of the excitation-energy spectra in $^{12}$B when gating on the 1022~keV peak in the $\gamma$-ray energy spectrum and three choices for the sideband, each with a width of 50 keV. The excitation-energy spectrum gated on the signal region ($E_\gamma=990\text{--}1054\, \mathrm{keV}$) of the 1022~keV $\gamma$-peak is depicted with a solid black line. The spectra corresponding to the different sideband regions are indicated with dashed and dotted lines in red, blue, green, and magenta.  The scaling factors ($k$) for the sideband ranges $E_\gamma = 1080\text{--}1130\, \mathrm{keV}$, 1130--1180~keV, and 1180--1230~keV are 1.45, 1.58, and 1.73, respectively. The entire sideband range $E_\gamma = 1080\text{--}1230\, \mathrm{keV}$, with a scaling factor of 0.53, is shown with dashed magenta lines. As the background shapes are the same for all choices of the sideband range within statistical uncertainties, the largest range was used in the further analysis. Similar procedures were performed for the 414~keV and 718~keV peaks, resulting in the choices for the sideband regions as shown in Fig.~\ref{fig:gamma_spectrum}.

Finally, the counts in the signal and sideband regions of the 414 and 1022~keV $\gamma$-peaks were extracted using a fit with Gaussian peaks second-order polynomial backgrounds. For the 718~keV peak, the fit included three Gaussians for the signal and a second-order polynomial for the background. The signal-to-background ratios for the 414, 718, and 1022~keV $\gamma$-rays were deduced to be 0.20, 0.40, and 1.11, respectively.

\begin{figure}[tbp]
\includegraphics[width=0.5\textwidth,clip]{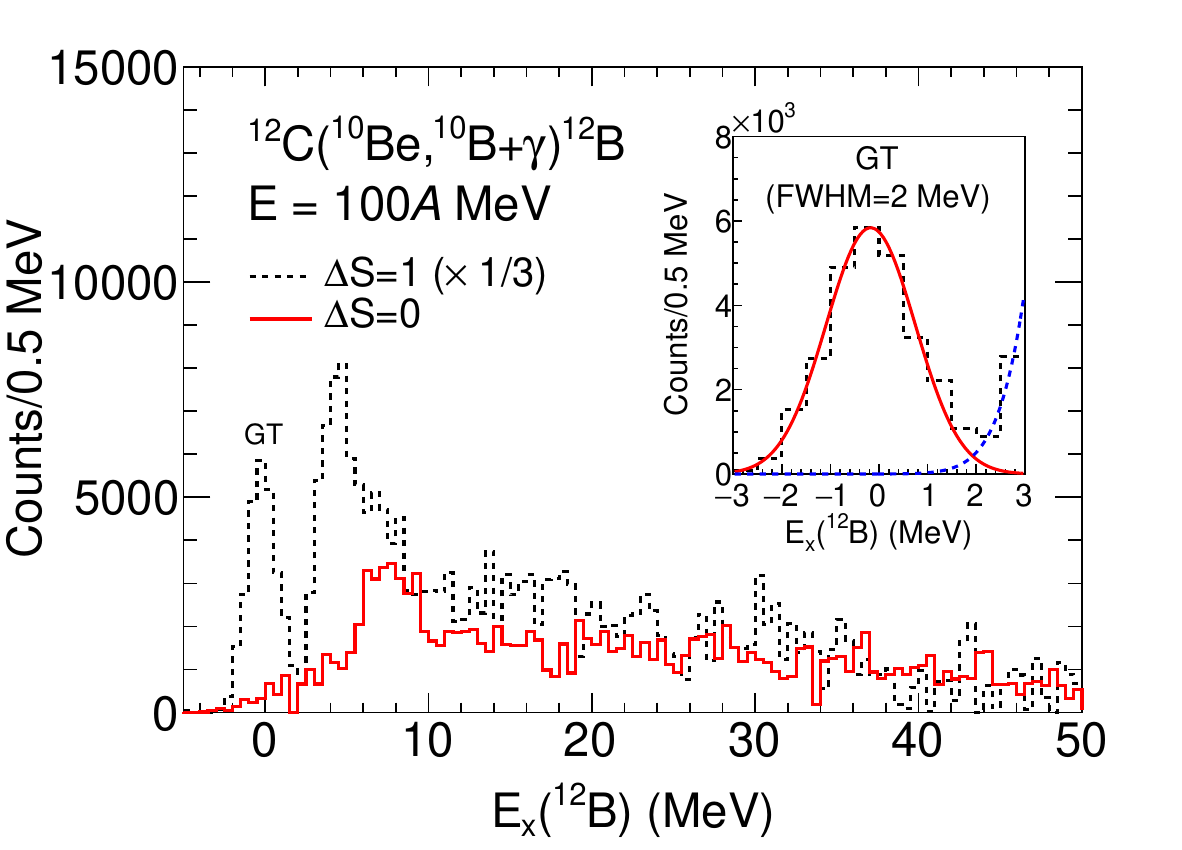}
\caption{Excitation-energy spectrum
of $^{12}$B for the $^{12}$C($^{10}\mathrm{Be}$,$^{10}\mathrm{B}$+$\gamma$) reaction gated on the $\Delta S=1$ (dashed black line) and $\Delta S=0$ (solid red line) filters. Inset shows the Gaussian fit of the Gamow-Teller (GT) peak with FWHM=2~MeV, which determines the excitation-energy resolution. }\label{fig:Ex12B}
\end{figure}

The excitation-energy spectra in $^{12}$B, reconstructed in a missing-mass calculation from the outgoing momenta of the $^{10}\mathrm{B}$ ejectiles detected in coincidence with the de-excitation $\gamma$-rays, are shown in Fig.~\ref{fig:Ex12B}. The black (red) histogram represents the excitation-energy spectrum with the $\Delta S=1$ ($\Delta S=0$) filter, after correcting for the feeding from the higher-lying states and subtracting the background. In the missing-mass calculation, the excitation of the 0.718 MeV ($\Delta S=1$) or 1.74 MeV ($\Delta S=0$) state was accounted for in the rest mass of the $^{10}\mathrm{B}$ ejectile.

A strong peak at $E_x=0$ MeV is observed in the $\Delta S=1$ spectrum, corresponding to the transition $^{12}\mathrm{C}(0^+, \text{g.s.})\rightarrow{}^{12}\mathrm{B}(1^+, \text{g.s.})$, which can only proceed with $\Delta S=1$. This peak has a width (FWHM) of 2~MeV, which determines the experimental resolution of the reconstructed excitation-energy as shown in Fig.~\ref{fig:Ex12B} inset.
When the $\Delta S=0$ filter is applied, this peak vanishes, showing the effectiveness in isolating the non-spin-transfer response. 

\begin{figure}[tbp]
\centering
\includegraphics[width=0.5\textwidth]{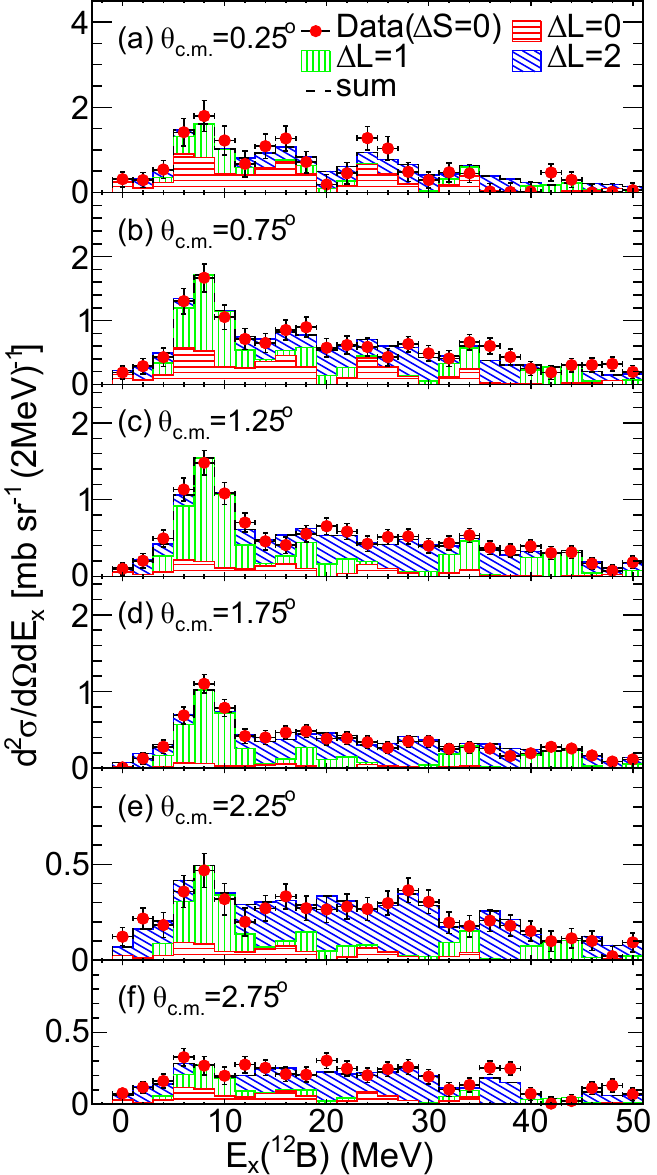}
\caption{Double-differential cross sections for the $^{12}$C($^{10}\mathrm{Be}$,$^{10}\mathrm{B}$$^\ast$[1.74 MeV])$^{12}$B reaction as a function of $^{12}$B excitation-energy, indicated with the red circles. The spectrum is plotted for various scattering angles $\theta_\textrm{c.m.}=0^\circ-3^\circ$.  The error bars on the data represent the statistical uncertainties. The colored histograms correspond to the different multipole contributions from the MDA.}  \label{fig:mda s0}
\end{figure}

\begin{figure}[tbp]
\centering
\includegraphics[width=0.5\textwidth]{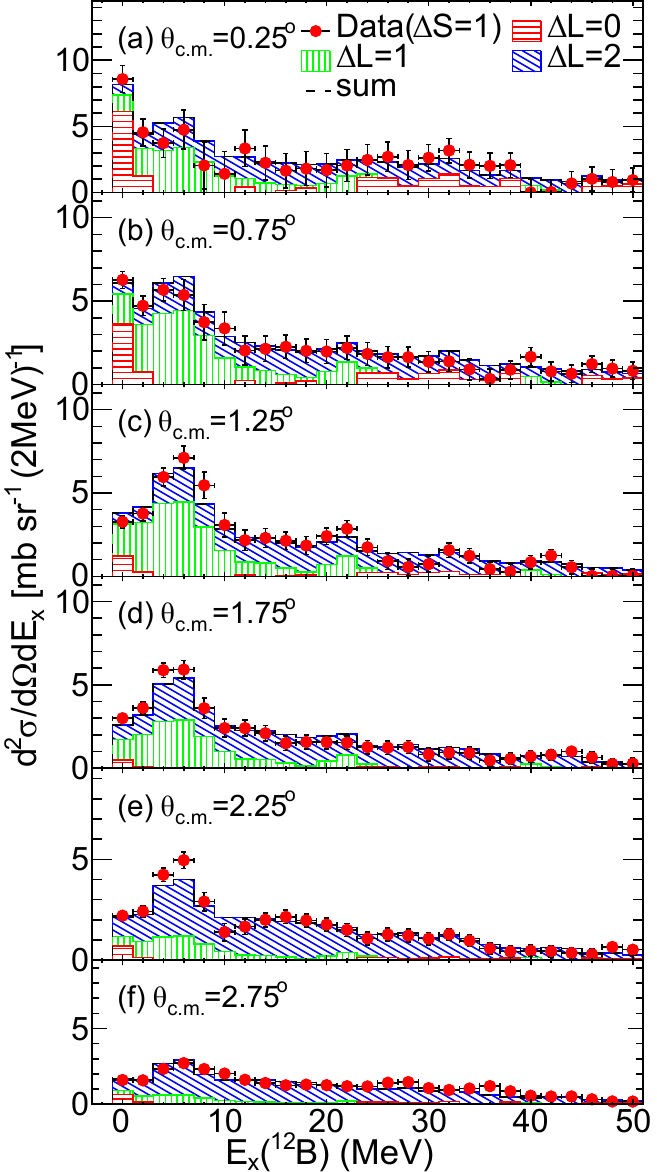}
\caption{Idem, but for the $^{12}$C($^{10}\mathrm{Be}$,$^{10}\mathrm{B}$$^\ast$[0.718 MeV])$^{12}$B reaction.}  \label{fig:mda s1}
\end{figure}

\subsection{Multipole decomposition analysis}
\begin{figure}[htbp]
\centering
\includegraphics[width=0.4\textwidth]{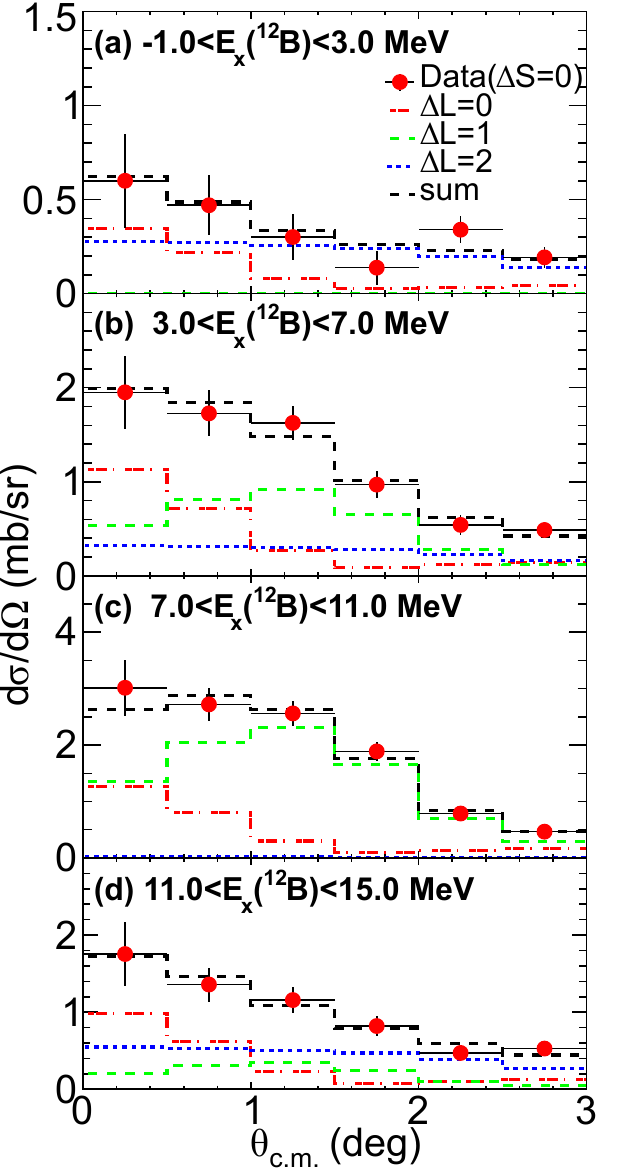}
\caption{The angular distributions for the $^{12}$C($^{10}\mathrm{Be}$,$^{10}\mathrm{B}$$^\ast$[1.74 MeV])$^{12}$B reaction for the $\Delta S = 0$ channel at different excitation-energy ranges are shown with the red data points. The results from the MDA using DWBA calculations for angular-momentum transfers of $\Delta L = 0$, $1$, and $2$ are represented by the colored lines, and their sum is shown with the black dashed line.}  \label{fig:xs s0}
\end{figure}

\begin{figure}[tbp]
\centering
\includegraphics[width=0.4\textwidth]{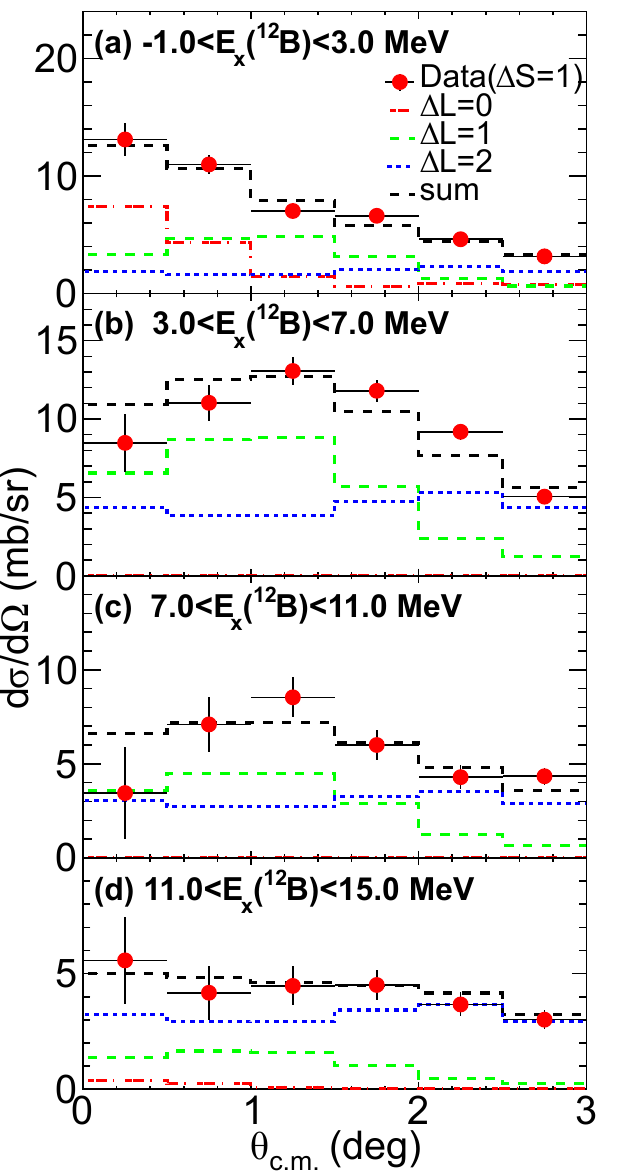}
\caption{Idem, but for the $^{12}$C($^{10}\mathrm{Be}$,$^{10}\mathrm{B}$$^\ast$[0.718 MeV])$^{12}$B reaction for the $\Delta S = 1$ channel.}  \label{fig:xs s1}
\end{figure}
The double-differential cross sections for the $^{12}$C($^{10}\mathrm{Be}$,$^{10}\mathrm{B}$+$\gamma$) reaction were determined from the counts obtained in the background-subtracted excitation-energy spectra. The cross sections were corrected for the acceptance of the S800 spectrometer, the efficiencies of the CRDCs, the efficiency of GRETINA for the coincident $\gamma$-rays, and the live-time of the data acquisition system. The measured double-differential cross sections as a function of $^{12}$B excitation-energy ($E_x$) for the $\Delta S=0$ and $\Delta S=1$ reactions at different scattering angles in the center-of-mass frame ($\theta_\textrm{c.m.}$) are shown with the red circles in Figs.~\ref{fig:mda s0} and \ref{fig:mda s1}, respectively.
The cross sections were binned into 2~MeV intervals in excitation energy, and the scattering angles were separated into 0.5$^\circ$ bins up to $3^\circ$. The systematic errors in the absolute normalization include uncertainties due to the background subtraction, the target thickness, and the $^{10}\mathrm{Be}$ beam intensity, with the latter being the dominant source at ${\approx}4\%$.

The double-differential cross sections shown in Figs.~\ref{fig:mda s0} and \ref{fig:mda s1} include excitations associated with various orbital angular momentum transfers ($\Delta L=0,1,2,\dotsc$). To disentangle the contributions from transitions with the different multipolarities, a multipole decomposition analysis (MDA)~\cite{BONIN1984,ICHIMURA2006} was carried out. In the MDA, the measured differential cross sections for each bin in $E_x$ were fitted using the least-squares method with a linear combination of angular distributions for different units of~$\Delta L$ calculated in Distorted-Wave Born Approximation (DWBA), i.e.,
\begin{align}
  \sigma ^\text{calc} (\theta_\text{c.m.},E_x)
  =\sum_{\Delta L} a_{\Delta L} \sigma ^\text{calc} _{\Delta L} (\theta_\text{c.m.},E_x)
\end{align}
where $a_{\Delta L}$ are the fitting parameters associated with each multipole contribution $\Delta L_i$ ($\Delta L=0,1,2,\dotsc$), all of which have non-negative values. 

The DWBA calculations were performed using the miscroscopic, double-folding code \textsc{fold/dwhi}~\cite{PETROVICH1977,Cook1984}. The effective nucleon-nucleon interaction of Love and Franey~\cite{Love-Franey1985} at 100~MeV was double-folded over the transition densities of the projectile-ejectile ($^{10}\mathrm{Be}$-$^{10}\mathrm{B}$) and target-residue ($^{12}$C–$^{12}$B) systems to construct the form factors. One-body transition densities (OBTDs) were calculated for the $^{10}\mathrm{Be}$-$^{10}\mathrm{B}$ and $^{12}$C–$^{12}$B systems with the shell-model code \textsc{NuShellX}@MSU~\cite{BROWN2014}. The optical model potential parameters (OPPs) for the entrance and exit channels were calculated using the methods described in Ref.~\cite{Tostevin2006}.
These calculations of OPPs utilize the double-folding model, assuming $^{12}$C and $^{12}$B densities calculated from spherical Hartree-Fock (HF) calculations using the SkX parameterization of the Skyrme interaction~\cite{brown1998new}. For $^{10}\mathrm{Be}$ and $^{10}\mathrm{B}$, Gaussian density distributions were used, with root-mean-square (rms) radii of 2.30~fm~\cite{ozawa2001measurements}. A Gaussian nucleon-nucleon (\textit{NN}) effective interaction, characterized by a range of 0.5~fm, was used~\cite{tostevin1999core}. The interaction strengths were derived from the tabulation provided in Ref.~\cite{ray1979proton}.

The angular distribution associated with monopole ($\Delta L = 0$) excitations peaks at $\theta _\text{c.m.} = 0^{\circ}$, whereas the angular distribution associated with dipole ($\Delta L = 1$) transitions peaks at about $\theta _\text{c.m.} \approxeq 1^{\circ}$ and that for quadrupole ($\Delta L = 2$) transitions is almost flat over the angular range up to about $\theta _\text{c.m.} \lesssim 3^{\circ}$. The MDA was thus performed with angular distributions $\Delta L = 0$, $1$, and $2$, as the inclusion of higher multipoles did not affect the fit. As a result, the $\Delta L=2$ component effectively incorporates possible contributions from higher multipolarities.
The calculated angular distributions were smeared with experimental angular resolutions before performing the MDA. 

The results from the MDA of the non-spin and spin-flip excitations in the $^{12}$B at various scattering angles are illustrated in Figs.~\ref{fig:mda s0} and \ref{fig:mda s1}, respectively. The stacked histograms in red, green, and blue represent the monopole ($\Delta L=0$), dipole ($\Delta L=1$), and higher multipole ($\Delta L \geqslant 2$) contributions, respectively. Figures~\ref{fig:xs s0} and \ref{fig:xs s1} display the fitted angular distributions used in the MDA across different excitation energy ranges for the $\Delta S=0$ and $\Delta S=1$ channels, respectively.

\section{Results and Discussions}~\label{results}
The well-known strong GT transition ($\Delta S=1$, $\Delta L=0$) from the ground state of ${}^{12}\mathrm{C}$ to the ground state of $^{12}$B with $B(\mathrm{GT}) = 0.99$~\cite{Kelley2017} was clearly observed, peaking at the forward angles as shown in Fig.~\ref{fig:mda s1} and Fig.~\ref{fig:xs s1}(a). The extracted monopole cross section for this transition in the first angular bin ($7.4\pm1.3$ mb/sr) in the latter figure is a factor of $1.95\pm0.34$ larger than expected based on the DWBA calculations with the shell-model inputs, after correcting for the fact that the measured GT strength for the transition from the $^{10}\mathrm{Be}$ ground state to the $1^{+}$ excited state in $^{10}\mathrm{B}$ at 0.718~MeV is 21\% smaller than obtained in the shell-model calculations. The GT strength for the transition to the $^{12}$B ground state is identical to the shell-model calculations, and no correction is required. This comparison is helpful for estimating event rates and planning future experiments using this probe.  

\begin{figure}[tbp]
\centering
\includegraphics[width=0.5\textwidth]{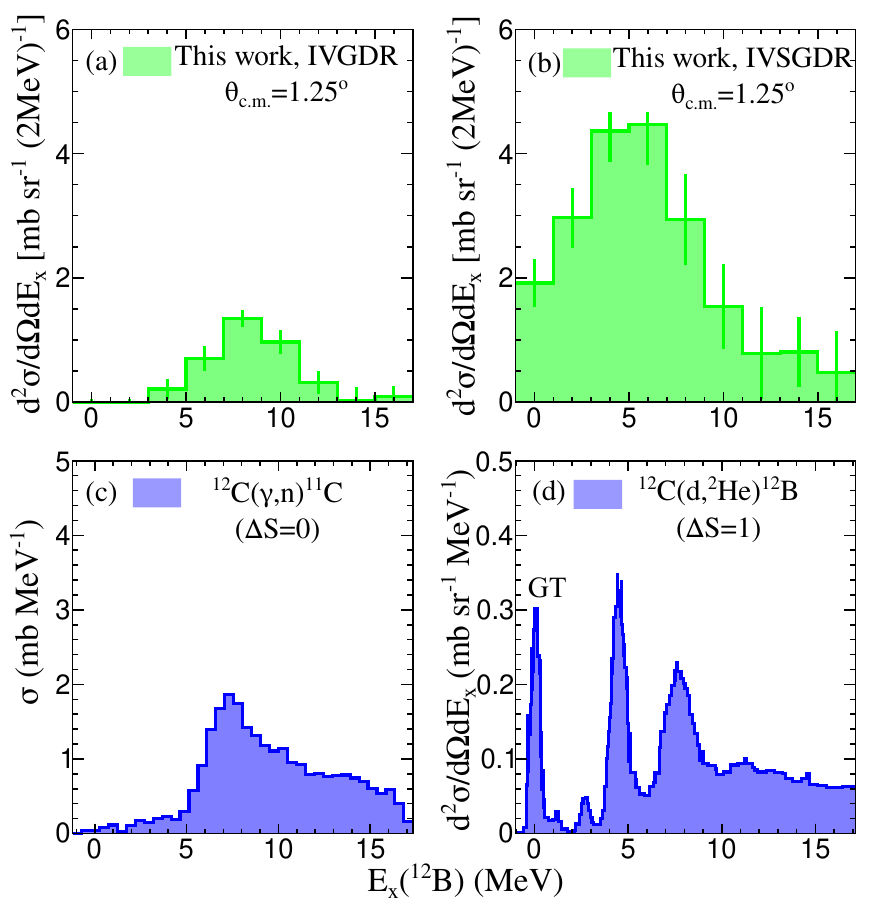}
\caption{(a) IVGDR and (b) IVSGDR cross sections extracted in the MDA at $\theta_\textrm{c.m.}=1.25^\circ$. (c) Photonuclear cross sections from the $^{12}\mathrm{C}(\gamma,n)^{11}\mathrm{C}$ reaction~\cite{AHRENS1975} selective to $\Delta S=0$ transition. (d) Cross sections for the $^{12}\mathrm{C}(d,{}^2\mathrm{He})^{12}\mathrm{B}$ reaction at $E_d = 270~\mathrm{MeV}$ ($\theta_\textrm{c.m.}=6^\circ$--$8^\circ$)~\cite{OKAMURA1994} selective to $\Delta S=1$ transition. See text for details.}  \label{fig:dipole}
\end{figure} 

The MDA results in Figs.~\ref{fig:mda s0} and \ref{fig:mda s1} also show strong dipole ($\Delta L=1$) excitations in both the $\Delta S=0$ and $\Delta S=1$ channels, peaking at $\theta_\textrm{c.m.}=1.25^\circ$, which are identified as the isovector giant dipole resonance (IVGDR, $\Delta S=0$ and $\Delta L=1$) and the isovector giant spin dipole resonance (IVSGDR, $\Delta S=0$ and $\Delta L=1$).
The double-differential cross sections of these resonances from the present work are shown in Fig.~\ref{fig:dipole}(a) and (b), respectively.
The IVGDR is strongly concentrated around $E_x \approx 7.5~\mathrm{MeV}$. 
For comparison, the photonuclear cross section from the $^{12}\mathrm{C}(\gamma,n)^{11}\mathrm{C}$ reaction~\cite{AHRENS1975,Yang1993}, which excites the dipole resonance, 
is also shown in Fig.~\ref{fig:dipole}(c). The IVGDR around $E_x = 7.5~\mathrm{MeV}$ was also observed in the $^{12}\mathrm{C}(n,p)^{12}\mathrm{B}$~\cite{OLSSON1993,Yang1993} and $^{12}\mathrm{C}(^{13}\mathrm{C},{}^{13}\mathrm{N})^{12}\mathrm{B}$~\cite{ICHIHARA1995} CE experiments, sensitive to both $\Delta S=0$ and $1$ excitations.  The differential cross sections of the present IVGDR data and the $(\gamma,n)$ cross section data show a similar shape peaking around 7.5~MeV.

Figure~\ref{fig:dipole}(d) shows the double differential cross section of the $^{12}\mathrm{C}(d,{}^2\mathrm{He})^{12}\mathrm{B}$ reaction at $E_d=270~\mathrm{MeV}$, $\theta_\textrm{c.m.}=6^\circ$--$8^\circ$~\cite{OKAMURA1994}. The $(d,{}^2\mathrm{He})$ probe is only selective to the $\Delta S=1$ channel. The two peaks at $E_x = 4.5$ and 7.5~MeV correspond to spin-dipole states, and the peak at $E_x=0~\mathrm{MeV}$ is the GT transition. The spin-dipole states at 4.5 and 7.5~MeV were also observed in the $^{12}\mathrm{C}(^{12}\mathrm{C},{}^{12}\mathrm{N})$ reaction at $E = 135~\mathrm{MeV}/u$~\cite{ICHIHARA1994PLB,Ichihara1994} and the $^{12}\mathrm{C}(^7\mathrm{Li},{}^7\mathrm{Be})$ reaction~\cite{Nakayama1991,NAKAYAMA1998, ANNAKKAGE1999}, which are selective to spin-transfer excitations. In the present IVSGDR distribution, a broad peak centered at $E_x=5~\mathrm{MeV}$ is observed in Fig.~\ref{fig:dipole}(b). Due to the limited experimental energy resolution, the two peaks at 4.5 and 7.5~MeV could not be separated in the data. The cross section observed for dipole transitions near $E_x = 0~\mathrm{MeV}$ is likely due to the excitation of the $2^-$ and $1^-$ states at $E_x = 1.67$ and 2.62~MeV in $^{12}\mathrm{B}$.

Interestingly, the MDA of the non-spin-transfer ($\Delta S=0$) data also reveals fragmented contributions from $\Delta L=0$ transitions, especially visible at $\theta_\textrm{c.m.}=0.25^\circ$, as shown in Fig.~\ref{fig:mda s0}(a). No Fermi transitions from the ${}^{12}\mathrm{C}$ ground state are expected, as its IAS is absent from ${}^{12}\mathrm{B}$. The only expected excitation that is associated with $\Delta L=0$ is the IVGMR, and its strength is likely highly fragmented in such a light system. The data suggest that an experiment with higher statistics might be worthwhile to study the IVGMR in $^{12}$B in more detail. Aside from the transition to the ground state, significant monopole contributions are absent in the spin-transfer data in Fig.~\ref{fig:mda s1}. We note that due to the lower signal-to-background ratio for the $\Delta S=1$ filter compared to the $\Delta S=0$ filter, the statistical uncertainties in the $\Delta S=1$ spectra are larger than those in the $\Delta S=0$ spectra, resulting in larger uncertainties in the MDA. In combination with the reduction of the solid angle and yield near $0^{\circ}$, the sensitivity for ($\Delta L=0$) strength in the $\Delta S=1$ analysis is lower than in the $\Delta S=0$ analysis. As concluded in Ref.~\cite{Scott2017}, increased beam intensities will be necessary to reduce the statistical uncertainties and increase the accuracy in the MDA.

\section{Summary and Outlook}
In summary, the $^{12}\mathrm{C}(^{10}\mathrm{Be},{}^{10}\mathrm{B}+\gamma)^{12}\mathrm{B}$ reaction was studied at $100 A~\mathrm{MeV}$ in the angular range $0^\circ\leqslant \theta_\textrm{c.m.}\leqslant 3^\circ$ at NSCL. It was demonstrated that by using this reaction it is possible to separately extract the isovector non-spin-transfer ($\Delta S=0$) and spin-transfer ($\Delta S=1$) responses in a single measurement by gating on $\gamma$-rays associated with the excitation and decay of the $0^{+}$ 1.74~MeV and $1^{+}$ 0.718~MeV states in $^{10}\mathrm{B}$, respectively. The extraction of the spin-transfer and non-spin-transfer responses only relies on known branching ratios for $\gamma$-decay from excited states in $^{10}\mathrm{B}$, and assumptions on the details of the reaction mechanism are not necessary. The ability for high-resolution Doppler reconstruction of the decay-in-flight $\gamma$-spectrum by using the GRETINA array is important for enhancing the signal-to-noise ratio when isolating the relevant $\gamma$-lines. The relatively long half-life of the 0.718~MeV level used for creating the spin-transfer filter reduces the associated signal-to-background ratio compared to that for the 1.74~MeV level used for creating the non-spin-transfer filter. However, the high GT strength of 3.51 for the transition to the 0.718 MeV level, allows for a meaningful extraction of the spin-transfer response.         

The double-differential cross sections were measured for an excitation energy range of up to 50~MeV in $^{12}$B in both the spin- and non-spin-transfer channels. A multipole decomposition analysis (MDA) was performed for the full excitation energy range to extract the contributions from transitions from $^{12}$C to $^{12}$B associated with different units of angular momentum transfer. The differential cross section for the transition to the $^{12}$B $1^{+}$ ground state was compared with DWBA calculations using shell-model inputs, and it was found that the theoretical cross section was a factor of $1.95\pm 0.34$ times lower than the experimental value. The MDA results revealed significant dipole contributions in both $\Delta S=0$ and $\Delta S=1$ channels, peaking at $\theta_\textrm{c.m.}=1.25^\circ$. The extracted distributions are consistent with previous CE experiments studying the isovector excitations in the $^{12}$C-$^{12}$B system, affirming the utility of the $(^{10}\mathrm{Be},{}^{10}\mathrm{B}+\gamma)$ probe in isolating spin-isospin excitations from the same experiment.

Although the model-independent extraction of spin-transfer and non-spin transfer responses in a single CE experiment is unique, the statistical uncertainties and experimental resolutions achieved in this experiment were larger than those obtained by using other CE probes. This is because a relatively low-intensity secondary $^{10}\mathrm{Be}$ beam was available, which necessitated the use of a thick reaction target to reduce statistical uncertainties at the cost of worsening the energy resolution. The $^{10}\mathrm{Be}$ beam intensities at next-generation rare-isotope beam facilities will exceed those used in the present work by up to a few orders of magnitude. This will also make it possible to study heavier nuclei and to reduce the target thicknesses to achieve better energy resolutions. Finally, the methods developed here can also be applied to the $(^{10}\mathrm{C},{}^{10}\mathrm{B}+\gamma)$ reaction, making it possible to extract the non-spin- and spin-transfer responses model-independently in both the $\Delta T_{z}=+1$ and $\Delta T_{z}=-1$ directions.     

\begin{acknowledgments}
We thank the staff at the National Superconducting Cyclotron Laboratory (NSCL) for their support during the experiment. This work was supported by the US NSF [PHY-2209429, PHY-1102511, PHY-1430152 (Joint Institute for Nuclear Astrophysics Center for the Evolution of the Elements), PHY-1068217, PHY-1404442, PHY-1419765, and PHY-1404343] and by the U.S. DOE, Office of Science, Office of Nuclear Physics, under Grant No. DE-SC0023633 (MSU).
GRETINA was funded by the U.S. DOE Office of Science. Operation of the array at NSCL was supported by NSF under Cooperative Agreement PHY-11-02511 (NSCL) and DOE under Grant No. DE-AC02-05CH11231 (Lawrence Berkeley National Laboratory). 
B. A. B. acknowledges support of the NSF grant PHY-2110365. 
U. G.  acknowledges support of the NSF grant PHY-2310059.
\end{acknowledgments}

\nocite{*}
\bibliography{ref}

\end{document}